\documentclass[twocolumn,showpacs,preprintnumbers,amsmath,aps,amssymb]{revtex4-1}
\usepackage{graphicx}
\usepackage{dcolumn}
\usepackage{bm}
\usepackage{color}

\begin{document}

\title{Thermodynamic geometry for a non-extensive ideal gas}

\author{J. L. L\'opez${}^{a}$} \email{jlopez87@uabc.edu.mx}
\author{O. Obreg\'on${}^{b}$} \email{octavio@fisica.ugto.mx}
\author{J. Torres-Arenas${}^{b}$} \email{jtorres@fisica.ugto.mx} 
\affiliation{${}^{a}$Departamento de Matem\'aticas, Facultad de Ciencias
  \\ Universidad Aut\'onoma de Baja California, A.P. 1880, C.P. 22860, Ensenada B. C. M\'exico \\}
\affiliation{${}^{b}$Divisi\'on de Ciencias e
Ingenier\'ias Campus Le\'on,
  \\ Universidad de Guanajuato, A.P. E-143, C.P. 37150, Le\'on, Guanajuato, M\'exico. \\
}

\begin{abstract}
 A generalized entropy arising in the context of superstatistics is obtained for an ideal gas. 
 The curvature scalar associated to the thermodynamic space generated by this modified entropy is calculated using two formalisms of 
the geometric approach to thermodynamics. 
Using the curvature/interaction hypothesis of the geometric approach to thermodynamic geometry it is found that 
as a consequence of considering a generalized statistics, 
an effective interaction arises but the interaction is not enough to give a phase transition.
This generalized entropy seems to be relevant in confinement or in systems with not so many degrees of freedom, so
it could be interesting to use such entropies to characterize the thermodynamics of small systems.
\end{abstract}

\maketitle

\section{Introduction}
Because of the existence of anomalous systems which do not seem to obey the rules of common statistics 
generally associated to non equilibrium processes, a more general statistics has been proposed 
based on superstatistics \cite{Obregon,Tsallis2,Beck} which considers large fluctuations of intensive quantities \cite{Beck}.    
We review here this formulation in which the intensive fluctuating quantity is the temperature. 
This fluctuation gives rise to a certain probability distribution characterized by a generalized Boltzmann factor. 
We can, in principle, associate an entropy for 
every probability distribution and in this context the Boltzmann-Gibbs entropy corresponds to the usual Boltzmann factor. It is, however,
possible to obtain other generalized expressions for entropies associated to different probability distributions \cite{Tsallis1} depending on one or 
several parameters \cite{Beck,Tsallis2}. In \cite{Obregon,Obregon1}, it was shown how to generate an entropy depending only on the probability.  
The particular entropy considered here arises from a generalized gamma distribution depending only on the probability $p_l$. 
This entropy has several interesting features and it seems to be relevant for particular thermodynamic systems
like confined systems \cite{Obregon1} and in this context we find an interesting and necessary application of such generalized entropies. 
The quantum version of this entropy which is a generalization of the Von Neumann entropy arises by means of a natural generalization 
of the replica trick \cite{Nana,Calabrese}.    

In the formalism of the geometric approach to thermodynamics, a geometric structure is given for usual thermodynamic systems by means of 
Riemannian geometry \cite{Weinhold,Rup,Rup1,Quevedo2,Quevedo1,Quevedo3}. Particularly in the context of the so called geometrothermodynamics \cite{Quevedo2,Quevedo3}, 
the geometrical relevant quantities, like the thermodynamic metric, are invariant with respect to Legendre transformations resembling the fact that 
the thermodynamic information does not depend on what fundamental relation (thermodynamic potential) is used.
In this formalism, the representation invariance of the metric and its corresponding curvature scalar has been proved for simple thermodynamic systems \cite{Quevedo3}.  
On the other hand, inspired in fluctuation theory, a distance between points in a thermodynamic space can also be defined \cite{Rup,Rup1,Rup2}, and we can 
associate to this space a thermodynamic metric, a corresponding Riemann tensor and consequently a curvature scalar $R$. 
Both approaches coincide in the physical interpretation of the curvature scalar as a 
manifestation of the existence of intermolecular interactions. When the curvature 
associated to the corespondent thermodynamic metric is non-zero, an interaction of some nature is present \cite{Rup1,Quevedo1}, this is known
as the curvature/interaction hypothesis. Other physical aspects that the scalar reveals, which has been proven for several thermodynamic systems,
is the existence of first order phase transitions. The curvature scalar diverges at some point if a phase transition exists. For some systems, the 
point where the scalar diverges happens to be the critical point where the phase transition occurs \cite{Rup2,Quevedo3}.  
In the thermodynamic geometry of fluctuation theory, the sign of the curvature scalar also provides additional information. For some systems it is clear that the sign of $R$
represents the kind of interaction, being attractive for a negative scalar and repulsive for a positive scalar \cite{Rup3}. The sign of $R$ can 
also be associated to the bosonic or fermionic nature of the thermodynamic system, the Bose and Fermi ideal gases are a clear example of this, 
we have $R < 0$ in the first case and $R > 0$ in the second case \cite{Janyszek}. For some other systems, it appears a change of sign in the curvature scalar, from negative 
to positive or the other way around. These cases generally appear where a different statistics, other than that of Boltzmann's, is considered \cite{Rup3}.
Following the vast literature of the subject related to the thermodynamic geometry of the two approaches considered here, we find that the sign 
interpretation is more clear in the formalism of G. Ruppeiner \cite{Rup3} but even so, it is not at all clear that this interpretation could be valid for 
all thermodynamic systems.
   
In this work we particularly consider the curvature of a non-extensive ideal gas characterized by a generalized non-extensive entropy.
The particular entropy we use depends only on the probability distribution and arises in the realm of 
superstatistics \cite{Obregon,Obregon1}. In order to get a better insight of the physics behind the curvature scalar of our thermodynamic system,
we calculate the curvature scalar using the two formalisms mentioned earlier, we will call these two scalars, 
the geometrothermodynamic scalar for the scalar calculated following the formalism followed in \cite{Quevedo1} and the fluctuation theory scalar, 
to the scalar calculated following \cite{Rup3}. We will find that the particular entropy (statistics) we propose \cite{Obregon,Obregon1} modifies the 
geometric structure of the generalized thermodynamic space considered, namely a generalized ideal gas, giving rise to the appearance of 
an effective interaction.
  
The paper is organized as follows: First in section II we explain how our modified entropy, and its associated Boltzmann factor, arises by assuming 
a particular probability distribution. In section III we briefly introduce first the formalism of geometrothermodynamics developed by H. Quevedo \cite{Quevedo1}
and describe how the thermodynamic metric is calculated. In this same section we also introduce the thermodynamic metric 
in the formalism of G. Ruppeiner \cite{Rup1}. We calculate the curvature scalar in both formalisms to further analyze and compare 
the thermodynamic information contained in the scalars using the interpretation of both formalisms.   
In section IV we make a discussion about the interpretation of both scalars and in section V we conclude and present the main results of our work.   

\section{Generalized entropies depending only on the probability distribution}
The Boltzmann factor depending on the energy $E$ of a microstate associated with a local cell of average temperature $1/\beta$ is given by

\begin{equation}
B(E) = \int f(\beta)e^{-\beta E} d \beta
\end{equation}   

\noindent and different distributions $f(\beta)$ lead to different Boltzmann factors. Following the procedure stated in \cite{Beck,Tsallis1} it is possible, in principle, to
associate a modified entropy to every Boltzmann factor. As an example we have that for the distribution $f(\beta) = \delta(\beta-\beta_0)$ the usual 
Boltzmann factor is recovered and from this, the Boltzmann-Gibbs entropy follows directly \cite{Tsallis1}. 
In \cite{Obregon,Obregon1} a Gamma distribution of the form

\begin{equation}
f_{p_l}(\beta) = \frac{1}{\beta_0 p_l \Gamma \left( \frac{1}{p_l} \right)} \left(  \frac{\beta}{\beta_0}\frac{1}{p_l}  \right)^{\frac{1-p_l}{p_l}}e^{-\beta/\beta_0 p_l}
\end{equation}  

\noindent was proposed where, by maximizing the appropriate information measure, the parameter $p_l$ can be identified with the probability and $\beta_0$ is the average inverse temperature. 
This distribution yields to the Boltzmann factor 

\begin{equation}
B_{p_l} (E) = (1+p_l\beta_0 E)^{-\frac{1}{p_l}}
\end{equation}   

\noindent that leads to the following generalized entropy \cite{Obregon1}

\begin{equation}
S = k\sum_{l = 1}^{\Omega} (1-p_{l}^{p_{l}}). 
\end{equation}

Associated to this generalized entropy there is a generalized $H$ function,
\begin{eqnarray}
H = \int d^3p \ \left(  e^{f \ln f} - 1 \right)
\end{eqnarray}

which it can be shown that it satisfies a generalized H-theorem \cite{Torres}. 
Using a Maxwell distribution to calculate this new $H$ function, keeping only the first order correction and the relation $H = - S/kV$, it follows

\begin{align}
S_{eff} =  & -kN\left[  \ln{ (n \lambda^3) } -\frac{3}{2} \right] \label{Sgen} \\ \nonumber 
& - \frac{kVn^2\lambda^3}{2^{5/2}} \left[ \ln^2{(n \lambda^3)}  -\frac{3}{2}\ln{(n \lambda^3)} +\frac{15}{16}  \right],
\end{align}

\noindent where $\lambda = \frac{h}{\sqrt{2\pi mkT}}$ can be identified with the mean thermal wavelenght, $k$ is the Boltzmann constant, 
$V$ is the volume and $T$ is the absolute temperature. The authors in \cite{Torres} study the thermodynamic properties of the corresponding generalized ideal gas. 
In this context, analysis of response functions shows a first correction having a universal form, that is, the same functional correction to 
all thermodynamic quantities derived from the generalized equations of state. 

In order to obtain a thermodynamic potential, we say that the conventional  linear relation between internal energy and temperature holds.
Within this approximation we obtain the following thermodynamic fundamental relation  

\begin{align}
S = & kN\ln{v} + \frac{3kN}{2}\ln{\frac{u}{b}} + \frac{3kN}{2} \label{GenEntropy} \\ \nonumber
& -\frac{kN}{2^{5/2}}\frac{b^{3/2}}{u^{3/2}v} \left[  \ln^2{ \left( \frac{b^{3/2}}{vu^{3/2}} \right)}  - \frac{3}{2}\ln{\left( \frac{b^{3/2}}{vu^{3/2}} \right)} +  \frac{15}{16} \right]
\end{align} 

\noindent where $u = \frac{U}{N}, v = \frac{V}{N}$ and $b = \frac{3h^2}{4\pi m}$. The first terms correspond to the usual entropy of the ideal gas, 
 $S = kN\ln{v} + \frac{3kN}{2}\ln{\frac{u}{b}} + \frac{3kN}{2}$. 
We notice that the Sackur-Tetrode expression for the entropy of the ideal gas $S = kN\ln{v} + \frac{3kN}{2}\ln{\frac{u}{b}} + \frac{5kN}{2}$ 
can be recovered by an ad-hoc fixing term as it was originally proposed by Gibbs. It is not possible to recover the $5kN/2$ term from the classical calculation we made 
but this does not affect the further analysis which involves derivatives of the entropy, and this constant term does not affect the final result.
At this point we have to clarify that the linear relation between 
internal energy and temperature that we have assumed makes our calculations to be more accurate for low densities or high temperatures and we will take this into account
to make the interpretation of the behavior of the curvature scalars.

A previous thermodynamic analysis was made corresponding to a system characterized by a particular interaction.  
The authors in \cite{Torres} considered a system of gas particles exposed to square-well and Lennard-Jonnes potentials.
These potentials are well defined and a respective Boltzmann factor $B(E) = e^{-\beta U}$ can be associated to these systems.
A further analysis with Monte Carlo simulations showed that when considering the generalized statistics and generalized entropy (\ref{GenEntropy}) 
the resulting thermodynamics can be modeled with a classical Boltzmann factor but with an effective interaction

 \begin{equation}
 B(E) = e^{-\beta U_{eff}},
 \end{equation}
 
 in the sense that the same form of the potential is recovered but with the redefinition of its parameters. 
 For the particular interactions considered and for particular thermodynamic states, the contribution generated by the new statistics 
 in the generalized potential turned out to be repulsive \cite{Torres} but the nature of the interaction depends in general on the thermodynamic states considered.  
 
 In the next section we move to the thermodynamic geometry formalism and calculate the scalar curvature $R$ associated to the fundamental relation (\ref{GenEntropy}). 
 It is our purpose to investigate and interpret the physical information contained in $R$ in two different approaches. We will calculate the geometrothermodynamic scalar 
 \cite{Quevedo1} and the fluctuation theory scalar \cite{Rup}. 
 The main reason to make the calculation in these two different formalisms 
 is because we are interested to see wether or not the two calculations show the same qualitative behavior in order to make an accurate physical interpretation
 according to these two, in principle, different theories. 
 The motivation and the origin of these two approaches is different but they have some common points of view regarding the physical interpretation of $R$
 in terms of its behavior.      

\section{Generalized thermodynamic metric and curvature}

As we have mentioned earlier, in this section we will calculate the thermodynamic metric and its corresponding curvature scalar associated to the entropy (\ref{GenEntropy})
in two formalisms; i) using geometrothermodynamics, namely using Quevedo's metric. In this formalisms we will refer to the curvature scalar as $R_G$. ii) using fluctuation theory, 
namely using Ruppeiner's metric. In this formalisms we will refer to the curvature scalar as $R_F.$

\subsection{Geometrothermodynamic approach}

In this formalism, the way in which the metric of the thermodynamic space is calculated makes it an object invariant under Legendre transformations.
This fact allows to include in the geometrical description of thermodynamics the invariance under thermodynamic potential representation \cite{Quevedo1,Quevedo3}. The 
thermodynamic potential is an essential part of the so called contact manifold which represents the thermodynamic space and its geometrical properties will be 
encoded in its related metric and geometrical objects related to it \cite{Quevedo1}. A contact manifold is characterized by three specific components 
$(\tau,\Theta,G)$, a $2n+1$-dimensional differential manifold $\tau$ with tangent space $T(\tau)$, a 
differential 1-form $\Theta$ on the cotangent manifold $T^{*}(\tau)$ such that an existing field of hyperplanes $\nu$ satisfies $\nu = \ker \Theta$ and a non degenerate 
metric $G$ on $\tau$ \cite{Quevedo1,Quevedo2}. We can choose a coordinate patch on $\tau$ $Z^A = \{\  \Phi,E^a,I^a \}\ $ with $a = 1,...,n$. In the particular thermodynamic 
representations $\Phi$ is identified with the thermodynamic fundamental relation, namely any thermodynamic potential containing all the thermodynamic information of the system. 
$E^a$ and $I^a$ correspond to the extensive and intensive variables, respectively. The general metric $G$ is given by

\begin{equation}
G = (d\Phi -I_adE^a)^2 + \Lambda(E_{a}I_a)^{2k+1}dE^{b}dI^{b}. \label{Metric1}
\end{equation}

\noindent where $k$ is an integer and $E_a = \delta_{ab}E^a,~I_a = \delta_{ab}I^a$. $\Lambda$ is an arbitrary Legendre invariant real function of the variables $E^a$ and $I^a$.
One special characteristic of thermodynamic systems is the fact that extensive and intensive variables are related among them through the equations of state.
The functional relation among these variables is given in a geometric way by means of a harmonic map 
$\varphi : \{\  E^a \}\ \rightarrow \{\ Z^A(E^a)  \}\ = \{\ \Phi (E^a),E^a,I^a(E^a) \}\ $. The space of coordinates $\mathcal{E}$ is a sub manifold of $\tau$ which is equipped 
with the non degenerate metric $G$, so the pullback of the harmonic map $\varphi^*$ induces a metric on $\mathcal{E}$, i. e.  $g = \varphi^*$. 
Given the equilibrium conditions given by the equations of state the simplest thermodynamic metric for the entropy representation in terms of 
internal energy $u$ and volume $v$ $s(u,v)$ is given by

\begin{align}
g = & \left[  \left( u\frac{\partial s}{\partial u}  \right)^{-1}\frac{\partial^2 s }{\partial u^2}du^2 
+ \left( v\frac{\partial s}{\partial v}  \right)^{-1}\frac{\partial^2 s }{\partial v^2}dv^2 \right]  \label{Metric2}  \\ \nonumber
& + \left[ \left( u \frac{\partial s}{\partial u}  \right)^{-1} + \left( v\frac{\partial s}{\partial v}  \right)^{-1}  \right] \frac{\partial^2 s}{\partial u \partial v}dudv, 
\end{align} 

which can be computed straightforwardly if the fundamental relation $s(u,v)$ is known.  We should remark that particular values of $\Lambda$ and $k$ has been chosen 
in order to get the metric (\ref{Metric2}). It is not clear that an arbitrary choice of $\Lambda$ and $k$ will display the same physical information in the curvature scalar and
It could also be possible that these kind of general metrics (\ref{Metric1}) will not be in general invariant under infinitesimal Legendre transformations \cite{Monsalvo}.
However, it has been shown that this metric is able to predict accurately the divergence of the curvature scalar at the critical point where 
the phase transition occurs for some ordinary thermodynamic systems \cite{Quevedo3}. In that sense, we will consider this metric as a good one to 
infer if our generalized system has some interaction or a phase transition, information 
that can be deduced from its corresponding curvature scalar. Nevertheless, we will compare the results with the curvature scalar using the metric of the fluctuation theory 
approach (\ref{RMetric1}) which was constructed with no reference to Legendre invariance at all \cite{Rup3}.  

We can compute now the components of the metric associated to the entropy (\ref{GenEntropy}), these are given by 

\begin{align}
& g_{uu} =- \frac{3}{2u^2} \left[ \frac{1-\frac{21\sqrt{2}}{256}x + \frac{17\sqrt{2}}{32}x\ln{ x } + \frac{5\sqrt{2}}{16}x \ln^2{ x }}{\frac{3}{2} - x \left(  \frac{27\sqrt{2}}{256}  -\frac{3\sqrt{2}}{32} \ln{x } -\frac{3\sqrt{2}}{16} \ln^2{x} \right) } \right], \label{QMetric} \\ \nonumber
& g_{uv} = g_{vu} = \frac{x}{uv}  \left[ \frac{\frac{3\sqrt{2}}{256} - \frac{15\sqrt{2}}{32}\ln{x } - \frac{3\sqrt{2}}{16} \ln^2{x}}{1 - x \left(  \frac{9\sqrt{2}}{128}  -\frac{\sqrt{2}}{16} \ln{x} -\frac{\sqrt{2}}{8} \ln^2{x} \right) } \right], \\ \nonumber
& g_{vv} =- \frac{1}{v^2} \left[ \frac{ 1-\frac{5\sqrt{2}}{64}x + \frac{3\sqrt{2}}{8}x\ln{x} + \frac{\sqrt{2}}{4}x \ln^2{x}}{ 1 - x \left(  \frac{9\sqrt{2}}{128}  -\frac{\sqrt{2}}{16} \ln{x} -\frac{\sqrt{2}}{8} \ln^2{x}      \right) } \right], \\ \nonumber
\end{align}

\noindent where we have introduced the dimensionless factor $x = \frac{b^{3/2}}{u^{3/2}v}$.  We notice that 
in the limit of low densities and high temperatures, that is when $x \rightarrow 0$, the entropy (\ref{GenEntropy}) and metric (\ref{QMetric}) approach respectively 
to those corresponding to the ideal gas \cite{Quevedo1,Quevedo3}. The curvature scalar for the metric (\ref{QMetric}) in terms of $x$ is then given by

\begin{align}
& R_G = -\frac{3 \cdot 2^{7}\left[  I_0 - 2^3 I_1 \ln{x} 
- 2^5 I_2 \ln^2{x} + 2^7 I_3 \ln^3{x} \right] }{G_0D(x)^3 } \label{QScalar} \\ \nonumber
& -\frac{3 \cdot 2^{7}\left[ 2^{8} I_4 \ln^4{x} - 2^{12}I_5\ln^5{x} -2^{14}I_6 \ln^6{x} \right]}{G_0D(x)^3} \\ \nonumber
& +\frac{3 \cdot 2^{7}\left[ 2^{16} I_7 \ln^7{x} - 2^{16} I_8 \ln^8{x} - 2^{19} I_9\ln^9{x} \right]}{ G_0D(x)^3} \\ \nonumber
& - \frac{3 \cdot 2^{7} \left[ 2^{21}I_{10} \ln^{10}{x} + 11 \cdot 2^{23} I_{11}\ln^{11}{x} - I_{12}\ln^{12}{x}  \right] }{ G_0D(x)^3 }, \\ \nonumber
& D(x) = (G_1 + G_2)\ln{x} + G_3\ln^2{x} + G_4\ln^3{x} + G_5\ln^4{x}.
\end{align}


\noindent where the $I_i's$, and $G_i's$ are dimensionless functions depending on $x$. We will analyze the behavior of this scalar 
 in the section of discussions where we will show the plot of $R_G$. In order to have a better insight with respect to the physical interpretation of the curvature scalar,
 in the next section we will also compute the curvature scalar in the fluctuation theory of thermodynamic geometry \cite{Rup,Rup1}.

\subsection{Fluctuation thermodynamic geometry approach}

Now we turn to the calculation in the formalism developed by G. Ruppeiner \cite{Rup,Rup1,Rup3}. The thermodynamic curvature in this formalism 
comes from fluctuation theory. The distance in the thermodynamic space arises from the calculation of the probability of a fluctuation of some variable 
$x^{\alpha}$ away from equilibrium of a system with fixed volume $V,$ using Einstein's formula \cite{Rup1}

\begin{equation}
\text{Probability} \propto \exp{\left( -\frac{V}{2}(\Delta l)^2 \right)},
\end{equation}

where the differential of distance squared $(\Delta l)^2$ defines a thermodynamic information metric $(\Delta l)^2 = g_{\alpha \beta} \Delta x^{\alpha} \Delta x^{\beta}$,
with $\Delta x^{\alpha} = (x^{\alpha} - x^{\alpha}_0)$, being $x^{\alpha}_0$ the value of the fluctuating variable $x^{\alpha}$ at equilibrium. 
The thermodynamic space
could be in general considered as $n$-dimensional and the $x^{\alpha}$ represent the independent fluctuating variables in this thermodynamic space.
Considering the fluctuation of two variables, $x^1$ and $x^2$ are two independent extensive variables or mechanical parameters.
The entropy can be considered to be a function of internal energy and the number of particles. With fixed volume, the variables 
of entropy $s(u,n) = S/V$ are the densities, $u = U/V$ $n=N/V$. The expansion to second order of the conditional probability allows to define the following metric

\begin{equation}
g_{\alpha \beta} = -\frac{1}{kV}\frac{\partial^2 s}{\partial x^{\alpha} \partial x^{\beta}}, \label{RMetric1}
\end{equation}    

where $k$ is Boltzmann's constant and $V$ is the fixed volume. We will need to rewrite the entropy (\ref{GenEntropy}) in terms of the densities $u,n$ in order to 
calculate the curvature scalar associated to (\ref{RMetric1}), the entropy is given by

\begin{align}
& S = kn \ln{\frac{1}{n}} + \frac{3kn}{2}\ln{\frac{u}{bn}} + \frac{3kn}{2} \label{REntropy} \\ \nonumber
& -\frac{kn}{2^{5/2}}\frac{b^{3/2}n^{5/2}}{u^{3/2}} \left[  \ln^2{ \left( \frac{b^{3/2}n^{5/2}}{u^{3/2}} \right)} - \frac{3}{2}\ln{\left( \frac{b^{3/2}n^{5/2}}{u^{3/2}} \right)} +  \frac{15}{16} \right], 
\end{align}

and the components of the metric following this formalism are given by

\begin{align}
& g_{uu} = \frac{3n}{2u^2V} \left[ 1- \frac{21\sqrt{2}}{256} z + \frac{17\sqrt{2}}{32}z \ln{z} + \frac{5\sqrt{2}}{16}z \ln^2{z}  \right],  \label{RMetric3} \\ \nonumber
& g_{nn} = \frac{5}{2nV} \left[ 1- \frac{23\sqrt{2}}{256} z + \frac{27\sqrt{2}}{32}z \ln{z} + \frac{7\sqrt{2}}{16}z\ln^2{z}  \right],  \\ \nonumber
& g_{un} = -\frac{3}{2uV} \left[1- \frac{23\sqrt{2}}{256} z + \frac{27\sqrt{2}}{32}z \ln{z} + \frac{7\sqrt{2}}{16}z \ln^2{z}  \right]. 
\end{align}   

where $g_{un} = g_{nu}$. In the last expression, we have used the dimensionless variable $z = \frac{b^{3/2}n^{5/2}}{u^{3/2}}$ which is convenient to measure how the system 
gets away from ideality. It is clear that in the limit $z \rightarrow 0$, the entropy (\ref{REntropy}) and metric (\ref{RMetric3})
tend to the entropy and metric corresponding to those of the ideal gas \cite{Rup1}.  With the metric (\ref{RMetric3}) the curvature scalar is given in this case by

\begin{align}
& R_F = \frac{5 \cdot 2^{10}z\left[  \bar{I}_0 - 2^7 \bar{I}_1 \ln{z} - 2^8 \bar{I}_2 \ln^2{z} \right]}{n D(z)^3} \label{RScalar1} \\ \nonumber
& +\frac{5 \cdot 2^{10}z\left[ - 2^9 \bar{I}_3 \ln^3{z} + 2^{9} \bar{I}_4 \ln^4{z} + 2^{14} \bar{I}_5\ln^5{z} \right]}{n D(z)^3} \\ \nonumber
& + \frac{ 5 \cdot 2^{10}z \left[ 2^{15}\bar{I}_6 \ln^6{z} + 7 \cdot 2^{17} \bar{I}_7 \ln^7{z} + 7^22^{16} \bar{I}_8 \ln^8{z} \right] }{n D(z)^3 }. \\ \nonumber
& D(z) = \bar{G}_0 + 2^4\bar{G}_1\ln{z} + 2^5 \bar{G}_2\ln^2{z} + \bar{G}_3 \ln^3{z}+ \bar{G}_4\ln^4{z} 
\end{align}

where $n = N/V$, so $R_F$ has dimensions of volume. Also here the $\bar{I}_i's$, and $\bar{G}_i's$ are dimensionless functions depending in this case on $z$. 
In the next section we will show the plots of these scalars and analyze 
them in terms of what their respective formalisms tell us about the interpretation of the 
behavior of $R$ as function of its independent variables. 

\section{Results and discussions}

Before we show the plots of the scalars (\ref{QScalar}) and (\ref{RScalar1}), we notice the following; $R_G$ is dimensionless and $R_F$ has dimensions of volume 
and we can see from the generalized entropy Eq. (\ref{GenEntropy}) that we have the natural parameters $m$ and $h$ involved in our expressions, so we will introduce the Bohr radius $a_0$
as a parameter of length in order to construct dimensionless quantities and express the plots of the curvature scalars in definite units. 
We can easily construct with $m,h,a_0$ the quantities $u_0 = \frac{h^2}{ma_0^2}$ of energy and $v_0 = a_0^3$ of volume. Using these quantities 
we construct and plot the reduced curvature scalars $R^{*}_{G}(u^{*},v^{*})$ and $R_F^{*}(u^{*},n^{*})$ where $u^{*} = u/u_0$, $v^{*} = v/v_0$, $n^{*} = nv_0$ and $R_{F}^{*} = R_F/v_0$.

In order to define the entropy (\ref{GenEntropy}) we have assumed a linear relation between internal energy and temperature, approximation which is 
valid in the region of low densities and high temperatures, so we are confident that our results are more accurate in these regions, that is when $x \rightarrow 0$ and $z \rightarrow 0$
which correspond to small deviations from the entropy of the ideal gas.
Following this argument, we have plotted the scalars as function of inverse of internal energy to show the regime of high temperatures 
near the origin as it is the region of validity of the approximation made. Figure 1 shows the reduced curvature scalar corresponding to Quevedo's metric $R^{*}_G$ 
in terms of inverse reduced internal energy. 
Figure 2 shows the reduced scalar $R^{*}_F$ corresponding to Ruppeiner's metric.
In both figures we fixed the reduced particle number density to $n^{*} = 10^{-4}$. A typical low density value was chosen as
$n = 10^{29}$, for which we have a reduced value of $n^{*} = 10^{-4}$.  
From these figures we see that both scalars consistently tend to the ideal gas scalar (zero scalar) 
in the limits $x \rightarrow 0$ and $z \rightarrow 0$. Figure 3 shows both scalars for a fixed lower density ($n^{*} = 10^{-5}$). 

\begin{figure}[h]
\centering
\includegraphics[width=7cm]{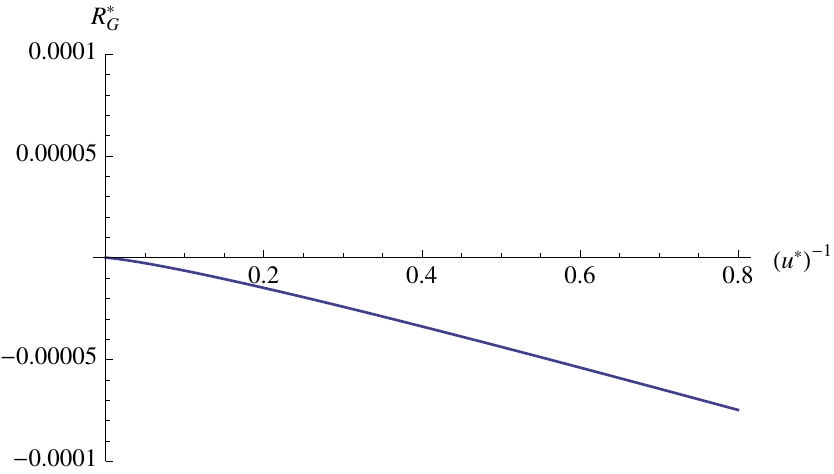}
\label{Figure1}
\caption{Reduced curvature geometrothermodynamic (GT) scalar $R^{*}_G$  as function of inverse of internal energy for fixed particle density $n^{*}=10^{-4}$. For high temperatures, or low values of $1/u^{*}$, the scalar 
consistently goes to zero. This limit corresponds to the limit $x \rightarrow 0$ which is the region where the approximations we made is valid.}
\end{figure}
  
\begin{figure}[h]
\centering
\includegraphics[width=7cm]{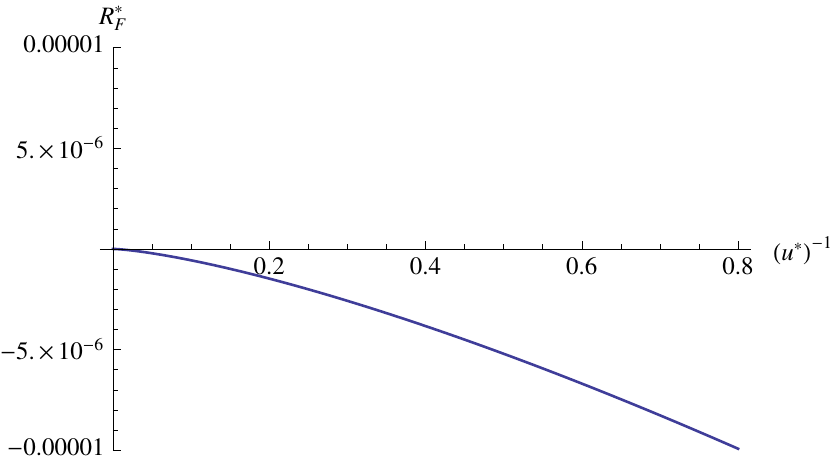}
\label{Figure2}
\caption{Reduced fluctuation theory (FT) curvature scalar $R^{*}_F$ as function of inverse of internal energy for fixed particle density $n^{*}=10^{-4}$.}
\end{figure}

\begin{figure}[h]
\centering
\includegraphics[width=7cm]{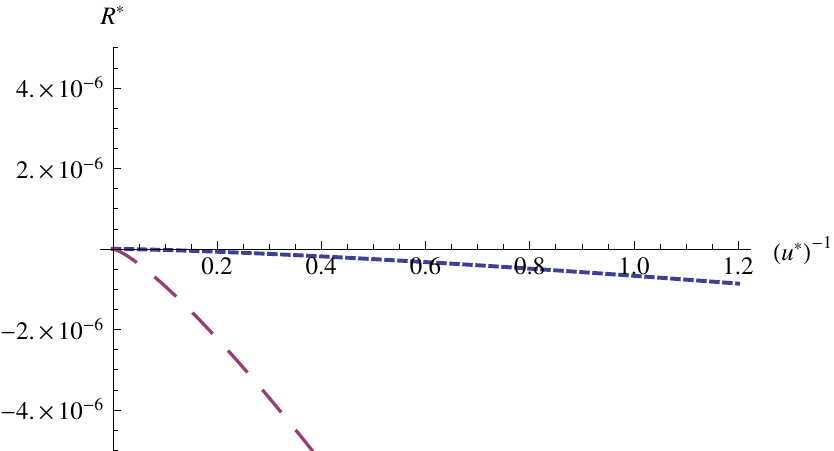}
\label{Figure3}
\caption{Simultaneous plots of $R^{*}_F$ (small dashed line) and $R^{*}_G$ (large dashed line) as function of inverse internal energy for fixed lower particle number density $n^{*} = 10^{-5}$.}
\end{figure}

As we have mentioned earlier, the interpretation of $R$ has its little differences in both approaches but they both agree in at least two things; First, they both coincide 
in the interpretation of the non-zero scalar as a manifestation of a non-vanishing intermolecular interaction; Second, the divergence of $R$ at some point
signs the existence of a phase transition. We also found in the literature that the sign of $R$ has a clear interpretation in the formalism 
of fluctuation theory and is less clear in the geometrothermodynamic approach. The fact that the sign of $R$ in Quevedo's formalisms does not clarify 
the kind of interaction is because the metric itself depends on an arbitrary global multiplicative function \cite{Quevedo1}. 
According to the results followed by Ruppeiner, the sign of $R_F$ shows the nature 
of the interaction, being attractive when $R < 0$ and repulsive when $R > 0$. In some cases, it reflects the fact that the system is composed of bosonic ($R< 0$)
or fermionic ($R>0$) particles, then some quantum aspects of the thermodynamic system are also contained in $R_F$ \cite{Janyszek}.
A generic discussion of the systems where the interpretation of the sign of $R$ is clear can be found in \cite{Rup3}.        

The scalars $R_F$ and $R_G$ of our thermodynamic system show a non-zero effective interaction and in the right limit, both consistently approach to the 
curvature scalar of the ideal gas, that is $R = 0.$ We have already clarified that a linear relation between internal energy and temperature has been assumed 
in the fundamental relations (\ref{GenEntropy}), (\ref{REntropy}) and in this expressions for the entropy, it is clear that in the respective limits $x \rightarrow 0$
and $z \rightarrow 0$, the generalized entropy tends to that of the ideal gas.  

\begin{figure}[h]
\centering
\includegraphics[width=7cm]{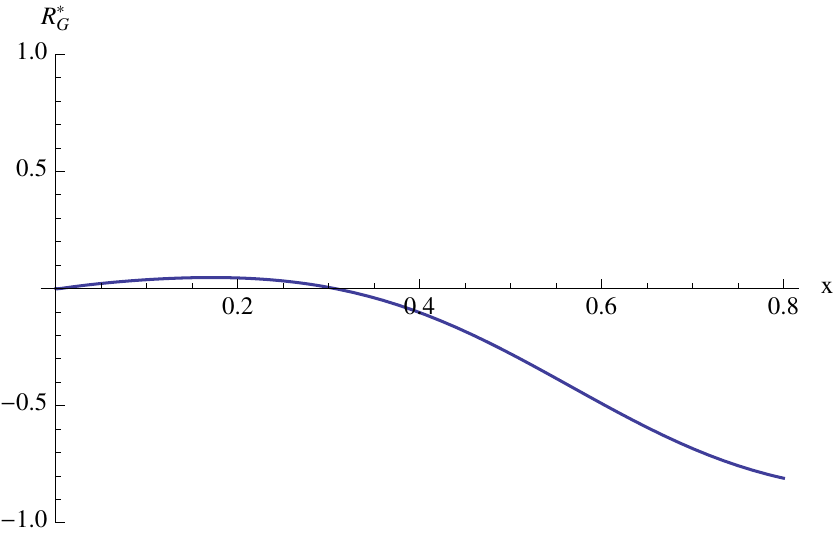}
\label{Figure3}
\caption{Dimensionless geometrothermodynamic curvature scalar $R^{*}_G$ as function of the parameter $x$.}
\end{figure}

\begin{figure}[h]
\centering
\includegraphics[width=7cm]{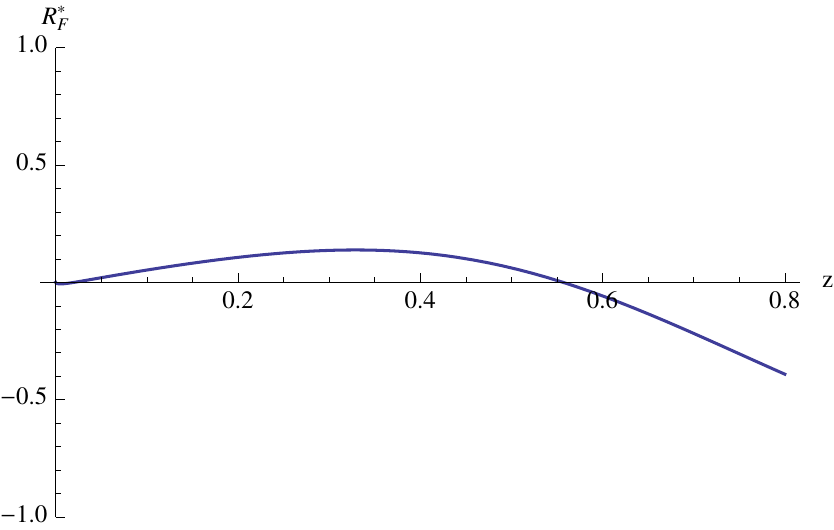}
\label{Figure3}
\caption{Dimensionless Fluctuation theory curvature scalar $R^{*}_F$ as function of the parameter $z$.}
\end{figure}

We show also in Fig. 4 and Fig. 5, the dimensionless scalars $R^{*}_G$ and $R^{*}_F$ in terms of the parameters 
$x$ and $z$ respectively. In this plots we can also see that the scalars have the right behavior in the ideal gas limits and here we notice that none of these scalars have 
divergences in the valid range of the approximation, they indeed do not diverge at any point in the whole range. This imply according to the interpretation 
of both formalisms that it does not exist a phase transition, so we have 
that an effective interaction appears as a consequence of assuming a different statistics, statistics which on the other hand defines the generalized 
non-extensive entropy of our non-extensive ideal gas.
  
\section{Conclusions}
 
We calculated the curvature scalar of a particular thermodynamic system, 
one that corresponds to the generalization of an ideal gas generated by a modified entropy. 
The modified entropy arises from a generalized Boltzmann factor (generalized statistics), namely a modified 
probability distribution. We found that such modification in the probability introduces an effective interaction.
Using the curvature/interaction hypothesis of thermodynamic geometry we found that  
such interaction shows up in a non-zero curvature scalar $R$.
Two different formalisms were used, the geometrothermodynamic 
approach and the fluctuation theory one. Both formalisms recover the limit 
of a conventional ideal gas characterized by a zero curvature scalar. 
Despite the non-zero value for the curvature $R$ implying a non-zero interaction, no evidence 
of a phase transition is obtained, namely the curvature $R$ does not diverge at any point.
It is remarkable the similarity in the results even when the metrics (\ref{Metric2}) and (\ref{RMetric1})
are in principle different.  

In the fluctuation theory approach, a formal interpretation to the sign of $R$ can be given describing the effective interaction, 
being attractive when $R < 0$ or repulsive when $R > 0$. We have therefore, according to Fig. 2, the appearance of an effective interaction
as a consequence of introducing a different statistics. Near the limit of the ideal gas, the effective interaction is attractive but it also shows regions where the interaction 
can become repulsive. 

A more rigorous statistical analysis is needed to better understand non-equilibrium inspired systems that obey a 
generalized statistic however, the curvature scalar 
is a useful tool revealing well defined characteristics of the thermodynamic system.       
     
\acknowledgments{ \noindent O. Obreg\'on was supported by CONACyT Projects No. 257919 and 258982, Promep and 
UG projects. J. Torres-Arenas acknowledge the University of Guanajuato for the Grant 740/2016 Convocatoria Institucional de Investigaci\'on 
Cient\'ifica and J. L. L\'opez was supported by CONACyT Grant 329847 and a PRODEP 
posdoctoral grant.}

\end{document}